\documentclass[conference]{IEEEtran}
\usepackage{amsmath, amssymb, graphicx, cite}
\usepackage{color}
\usepackage{courier}
\usepackage{url}
\usepackage{caption}

\providecommand{\mc}[1]{\mathcal{#1}}
\newcommand{\Real}{\mathbb{R}}
\newcommand\Mark[1]{\textsuperscript#1}

\begin{document}
\title{Computing Scalable Multivariate Glocal Invariants of Large \ (Brain-) Graphs}

\author{
    \IEEEauthorblockN{
Disa Mhembere\Mark{1}, %
William Gray Roncal\Mark{1}\textsuperscript{,}\Mark{2}, 
Daniel Sussman\Mark{1},} 

\IEEEauthorblockN{
Carey E.\ Priebe\Mark{1}, 
Rex Jung\Mark{3}, 
Sephira Ryman\Mark{3}, 
R.\ Jacob Vogelstein\Mark{2},
Joshua T.\ Vogelstein\Mark{1}\textsuperscript{,}\Mark{4}\textsuperscript{,}\Mark{5}, 
Randal Burns\Mark{1}} 
   
   \IEEEauthorblockA{
\Mark{1}Johns Hopkins University, 
\Mark{2}Johns Hopkins University Applied Physics Laboratory, }
   \IEEEauthorblockA{
\Mark{3}University of New Mexico,
\Mark{4}Duke University,
\Mark{5}Child Mind Institute
}
\vspace{-25pt}    
}
\maketitle
\newpage
\setcounter{page}{1}

\vspace{-15pt}
\begin{abstract}
Graphs are quickly emerging as a leading abstraction for the representation of data.
One important application domain originates from an emerging discipline called ``connectomics''. Connectomics studies the brain as a graph; vertices correspond to neurons (or collections thereof) and edges correspond to structural or functional connections between them. To explore the variability of connectomes---to address both basic science questions regarding the structure of the brain, and medical health questions about psychiatry and neurology---one can study the topological properties of these brain-graphs. We define multivariate glocal graph invariants: these are features of the graph that capture various local and global topological properties of the graphs.  We show that the collection of features can collectively be computed via a combination of daisy-chaining, sparse matrix representation and computations, and efficient approximations.   
Our  custom open-source \textit{Python} package  serves as a back-end to a Web-service that we have created to enable researchers to upload graphs, and download the corresponding invariants in a number of different formats. 
Moreover, we built this package to support distributed processing on multicore machines.
This is therefore an enabling technology for network science, lowering the barrier of entry by providing tools to biologists and analysts who otherwise lack these capabilities. As a demonstration, we run our code on 120 brain-graphs, each with approximately 16M vertices and up to 90M edges.
\end{abstract}

\section{Introduction}

A wide range of naturally occurring phenomena can be accurately depicted as graphs. Subsequently, graph visualization and analysis is of ubiquitous interest in industry and academia alike---with many applications such as social network analysis \cite{Bhaumik2012}, and recently, human brain mapping \cite{Craddock2013}. Functional and diffusion Magnetic Resonance Imaging (MRI) techniques have proven to be valuable tools for the creation of high-resolution brain-graphs\cite{Hagmann2008}, referred to as \textit{connectomes}. These brain-graphs have great potential to unlock physiological, functional, and structural unknowns within the human brain; thereby advancing fields of study like psychiatry and neurology, by extracting biologically-relevant characterizations. Utilizing brain-graphs as biomarkers requires extracting information from the graphs that is potentially informative with regard to the covariates of interest.  However, to date, no biomarkers have been useful for clinical diagnoses for any psychiatric disease category \cite{APA}.
We hypothesize this lack of efficacy of contemporary biomarkers may be due to two factors.  First, most analyses operate on region-wise graphs, rather than voxel-wise graphs, which reduces the number of vertices from $\mathcal{O}(10^7)$ to $\mathcal{O}(10^2)$.  This is a substantial dimensionality reduction which almost certainly imposes severe biases and discards valuable information.  Second, even given these relatively small graphs, the  graph-derived features typically used are relatively simple properties of the graphs.  The chosen features, therefore, may further discard clinically useful information.  The reduction is largely due to computational reasons: computing graph features can be computationally daunting, exact computations often being super-exponential in the number of vertices.   Existing frameworks such as igraph \cite{Csardi2006}, networkx \cite{Hagberg2008}, and BCT \cite{Rubinov2010} do not scale well to large graphs. 

To address these computational deficiencies, we formally define multivariate glocal graph invariants. These graph-derived features capture a variety of local and global properties of the graphs. By utilizing 
sparse matrix representations and computations, in conjunction with 
certain approximations, we can daisy-chain computations to efficiently compute all of these glocal invariants on large graphs.  All our code is implemented in an open-source \emph{Python} package.

Additionally, we provide Web-services with both programmatic and point-and-click interfaces to enable investigators or analysts who lack graph analytics expertise or resources to benefit from graph processing. 

Finally, we provide modules to build and compute invariants specifically for connectivity brain-graphs (i.e. connectomes) given fiber tractography streamline input data. 

\section{Multivariate Invariants}

Computing graph invariants provides a uniform platform upon which network connectivity across graphs of varying composition may be analyzed.  Let $\mc{G}$ be the set of all graphs, where $G=(\mc{V}_G,\mc{E}_G) \in \mc{G}$, and $\mc{V}_G$ is the set of  $n$ vertices $v$ for graph $G$, and $\mc{E}_G=\{u \sim v : u,v \in \mc{V}_G\}$ is the set of $m$ edges amongst $\mc{V}$. Let $A=(a_{uv})$ and $A'=(a_{uv}')$ correspond to the adjacency matrix representation for graphs $G$ and $G'$, respectively; that is $a_{uv}=1$ if and only if $u \sim v \in \mc{E}_G$. Let $G,G' \in \mc{G}$ be isomorphic to one another whenever $\exists$ such a $\pi:a_{uv} = a_{\pi(u)\pi(v)}'$ for all $u,v \in \mc{V}$; where $\pi : \mc{V} \to \mc{V}$ is a permutation function (bijection). Let $G=\pi(G')$ denote that $G$ and $G'$ are isomorphic to one another. Let $l(u,v)$ be the minimum number of edges required to traverse between vertices $u, v \in \mc{V}$. Let the $j$-hop \emph{neighborhood} of a graph $G$ around vertex $v$ be denoted by $N_j[v;G]$, where $N_j[v;G] = \{u \in \mc{V}:l(u,v) \le j\}$. Let $\Omega(\mc{V}')$ denote the induced subgraph of $\mc{V}' \subseteq \mc{V}$, that is the graph containing vertices in $\mc{V}'$ and all edges amongst them.  Let $size(G)=m$ denote the number of edges in the graph, and let $[n]=\{1, 2, \ldots, n\}$.

A \emph{global} invariant is a function of a graph that is invariant to permutations, that is, $\Phi = \{\phi : \mc{G} \to \Real^d$ s.t. $\phi(G)=\phi(G')$ whenever $G = \pi(G')\}$. Examples include number of vertices and edges, as well as max-degree, average path length, etc.    A \emph{local} (vertex-based) invariant is a function of a graph indexed by a vertex that is invariant to local permutations, $\Psi_j = \{ \psi_j : \mc{G} \times \mc{V} \to \Real^p$ s.t. $\psi(\Omega(N_j[v;G]))=\psi(\Omega(N_j[v;G']))$.  For example, the degree of a vertex is a local invariant.
A \emph{glocal} invariant is the collection of local invariants for all vertices, $\Xi_j=\{\xi_j : \mc{G} \to \Real^{d \times n}$ s.t. each $n$-dimensional vector is a local invariant as defined above$\}$.  We will give several examples below. 
The invariants we compute are primarily selected due to their utility in revealing underlying network features \cite{Pao2011}.  They are arranged to vaguely reflect the degree of topological complexity they incorporate.

\vspace{2 mm}
\noindent\textbf{Degree Vector} Deg $\in [n]^n$
is an $n$-dimensional vector where each element is an integer less than or equal to the number of vertices $n$.  The degree of a vertex is defined as the number of edges incident to it, so we can compute Deg via a simple matrix vector multiply operation: Deg$(v,G)=A \mathbf{1}$, where $\mathbf{1}$ is an $n$-dimensional vector of ones. 

\vspace{2 mm}
\noindent\textbf{Scan Statistic-$i$ Vector} SS-$i$ $\in [n]^n$
 is another $n$-dimensional vector where each element is the scan statistic of a particular vertex \cite{Pao2011}.  A vertex scan statistic counts the number of edges in the subgraph induced by  its $i$-hop neighborhood: $size\left(\Omega(N_i[\upsilon; G])\right)$.  SS-$j$ is the vector of these local invariants.  These can be computed in an embarrassingly parallel fashion using only very local graph properties whenever $i$ is small enough.  For the brain-graphs of interest, the graphs are sufficiently connected such that we can only efficiently compute SS-1 at this time.  

\vspace{2 mm}
\noindent\textbf{Number of Local $3$-Cliques Vector} NL-3 $\in [\binom{n}{3}]^n$ While Deg and SS-1 only consider pairwise interactions, NL-$j$ considers $j$-way interactions.   
NL-$j$ counts the number of $j$-cliques  in which each vertex participates. Tsourakakis \cite{Tsourakakis2008} demonstrated that an eigendecomposition of the adjacency matrix of the graph can be used to compute NL-$j$ for $j=3$ (also called triangles).  This approximation, however, does not work with modification for $4$-cliques. Thus, although number of local $j$-cliques would be an interesting multivariate glocal invariant, we could not design an efficient algorithm to approximate it at this time.  For NL-3 however, in experiments, we achieved 99.9\% accuracy  on graphs with 3,000 vertices and $\sim$4 million edges, using only 1 eigen-pair. The main computation is NL-3$(v,G) \approx  \frac{1}{2} \sum_{k=1}^{K} \lambda^3_k(G) x^2_{vk}(G) $ where, $x_{vk}(G)$ is the $v^{th}$ entry of the $k^{th}$ eigenvector {of $A$}, and $\lambda_k(G)$ is the $k^{th}$ eigenvalue, $K$ is the number of eigen-pairs used in the approximation.  

\vspace{2 mm}
\noindent\textbf{Clustering Coefficient Vector} CC $\in [\binom{n}{3}]^n$ combines Deg and NL-3 to assess the  relative amount of connectivity as compared to the potential total amount of connectivity for each vertex. To efficiently compute CC we utilize the previous computations, thus obtaining this additional glocal invariant is essentially free \cite{Saramaki2007}.  For each vertex, we compute $CC(v,G) = 2 \times NL$-$3(v,G)/({Deg(v,G)(Deg(v,G)-1)})$.

\vspace{2 mm}
\noindent\textbf{Latent Position Matrix} LP-$k$ $\in \Real^{k \times n}$ is a $k$-dimensional estimate of the latent positions of each vertex \cite{Hoff02}.   Latent position random graph models are elegant statistical models of random graphs with many desirable properties. 
In short, associated with each vertex is a latent position vector $x_v \in \Real^k$, and the probability of an edge between vertex $u$ and $v$ is a function of the dissimilarity between the latent positions $\kappa(x_u,x_v)$.
It has been shown that an eigendecomposition of the adjacency matrix of a graph yields universally consistent estimators for various parameters of a certain general class of latent position models \cite{Sussman2012a}.    Thus, this multivariate glocal invariant utilizes global topological properties, rather than just $j$-neighborhoods as in the previous glocal invariants.  For efficiency, we calculate the first $k$ eigenvectors via  the Lanczos  algorithm \cite{Lanczos1950}.  Moreover, since we estimate NL-3 via the eigendecomposition, our NL-3 approximation becomes essentially free after computing LP-1.  

Thus, the total collection of all multivariate glocal graph invariants are available after an eigendecomposition, a couple of local searches, and a few matrix vector multiply operations.  

\section{Software Tools}
We have developed novel tools to compute multivariate invariants on large scale graphs, including voxel-wise brain-graphs. The tools we have developed are designed to run on the CPU with as little as one core and a minimum of 8 GB of RAM. These software tools can be publicly accessed via a \textit{stand-alone Python Package} or our \textit{Web-Services}.

\subsection{Stand-alone Package}
The package, written in \textit{Python 2.7}, provides users with the ability to download, utilize and freely modify the existing implementations of invariant calculations. Additionally, we developed a set of command-line tools (CLTs), via executable scripts; these provide a high-level and simple way to safely interact with the actual data processing scripts. These CLTs are appropriate for, and have been successfully deployed \cite{Roncal2013a} in, distributed environments for the parallel processing of graphs. Finally, we provide scripts to build MRI fiber streamline graphs. This requires two inputs: 
(i) a fiber streamline file, in MRI Studio DAT format, and 
(ii)	a mask composed of two files that together describe the regions of interest (ROIs), in XML and RAW format. Any voxels outside these ROIs are excluded when creating graph edges.

The resulting brain-graph is stored as a sparse, compressed, and symmetric square matrix. We further elaborate on the methodology for building these connectome graphs in Section \ref{Experiments}. This open-source package is downloadable from  \texttt{\url{https://github.com/openconnectome/MR-connectome/tree/stand-alone}}. The distribution also includes several example scripts.

\subsection{Web-services}

Our Web-services are accessible at \texttt{\url{http://openconnecto.me/graph-services/}}, run on our own data-intensive cluster. These Web-services can be invoked through a graphical user interface via web browser, or programmatically via the command line. Both modes permit multiple-subject jobs, enabling the processing of several graphs with a single action. Programmatic interface functionality is well documented on the website with several example \textit{Python} scripts and command line calls to each service. The Web-services actively use our stand-alone package as the data processing back-end for graph building and invariant calculations. Available Web-services include:

(i) computing invariants on dense or compressed sparse column (CSC) square matrices in \textit{MATLAB} (MAT) format, 
(ii) conversion of invariant and graph files between MAT, NPY and CSV formats,
and (iii) building connectome graphs from diffusion MRI data.

\section{Experiments}\label{Experiments}

To demonstrate how our software can be used, we programmatically computed the invariants of 120 subjects' connectome graphs using our Web-services (see Figure \ref{fig:comp_pipe}).

\subsection{Computational Methodology}
To generate the brain-graphs, we begin with derivative data in the form of fiber tractography streamlines and a mask. These fiber streamlines, and the mask are by-products of MIGRAINE, a pipeline for efficiently computing large brain-graphs from raw multimodal magnetic resonance imaging data \cite{Roncal2013a}. MIGRAINE ingests the raw structural and diffusion MRI scans. The mask---which may describe ROIs, brain masks or grey matter---is applied to the data to exclude voxels outside the masked region from the brain-graph. Based on fiber tracing, we allocate edges to each pair of remaining voxels that are connected by fibers. The count of fibers between a pair of connected voxels yields a weight for each graph edge. These edge weights are later thresholded and binarized because we compute only unweighted invariants.

\begin{figure}[!ht]
\centering \resizebox{0.45\textwidth}{!}{\includegraphics{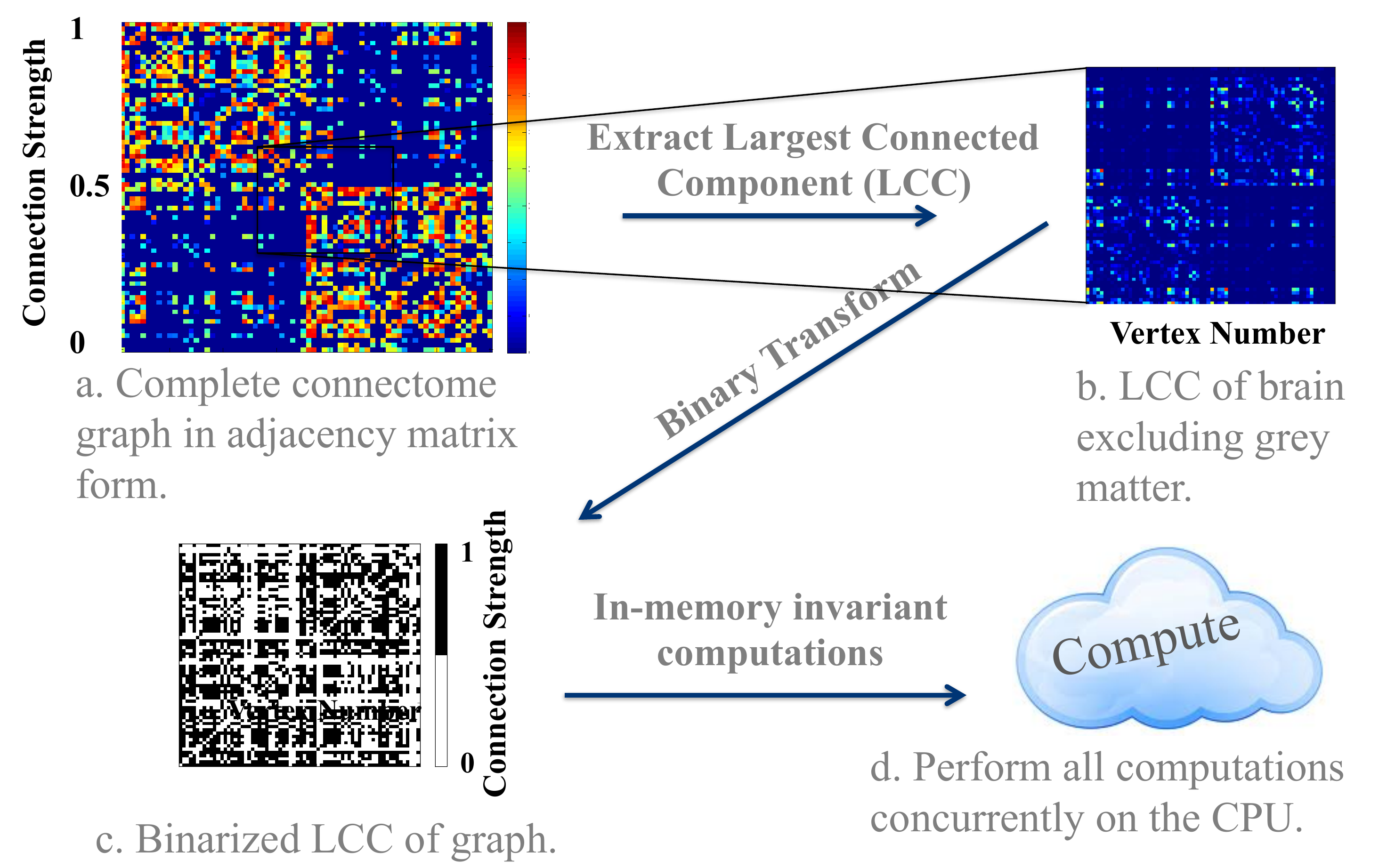}}
\vspace{-5pt}
\caption{Pipeline showing data transformation phases where we extract the LCC prior to brain-graph binarization and invariant computation.}  \label{fig:comp_pipe}
\vspace{-5pt}
\end{figure}

As depicted in Figure {\ref{fig:comp_pipe}} we extract a subset of the full brain-graph by taking the Largest Connected Component (LCC) \cite{Jones2001}. Our implementation includes a flag that enables operating on the LCC for the following reasons:

\begin{itemize}
	\item The LCC is by definition a connected graph, thus eigenvector embeddings of it are especially interesting due to their theoretical properties \cite{Sussman2012}, \cite{Fishkind2012}.%

	\item The LCC is of particular interest in many applications---thus computed invariants on the LCC are useful to a wide variety of researchers \cite{Zalesky2012}.%
	\item The LCC discards isolates and other vertices that are potentially ``noisy''. 
	Graph size is reduced from $O(10^7)$ to $(10^5)$ vertices while graph order is marginally affected with $>$99\% of edges remaining. This substantially decreases processing time by 94\% to under 2 hours per graph. 
\end{itemize}

\subsection{Results}

\begin{figure}[!ht]
\vspace{-8pt}
\centering \resizebox{0.5\textwidth}{!}{\includegraphics{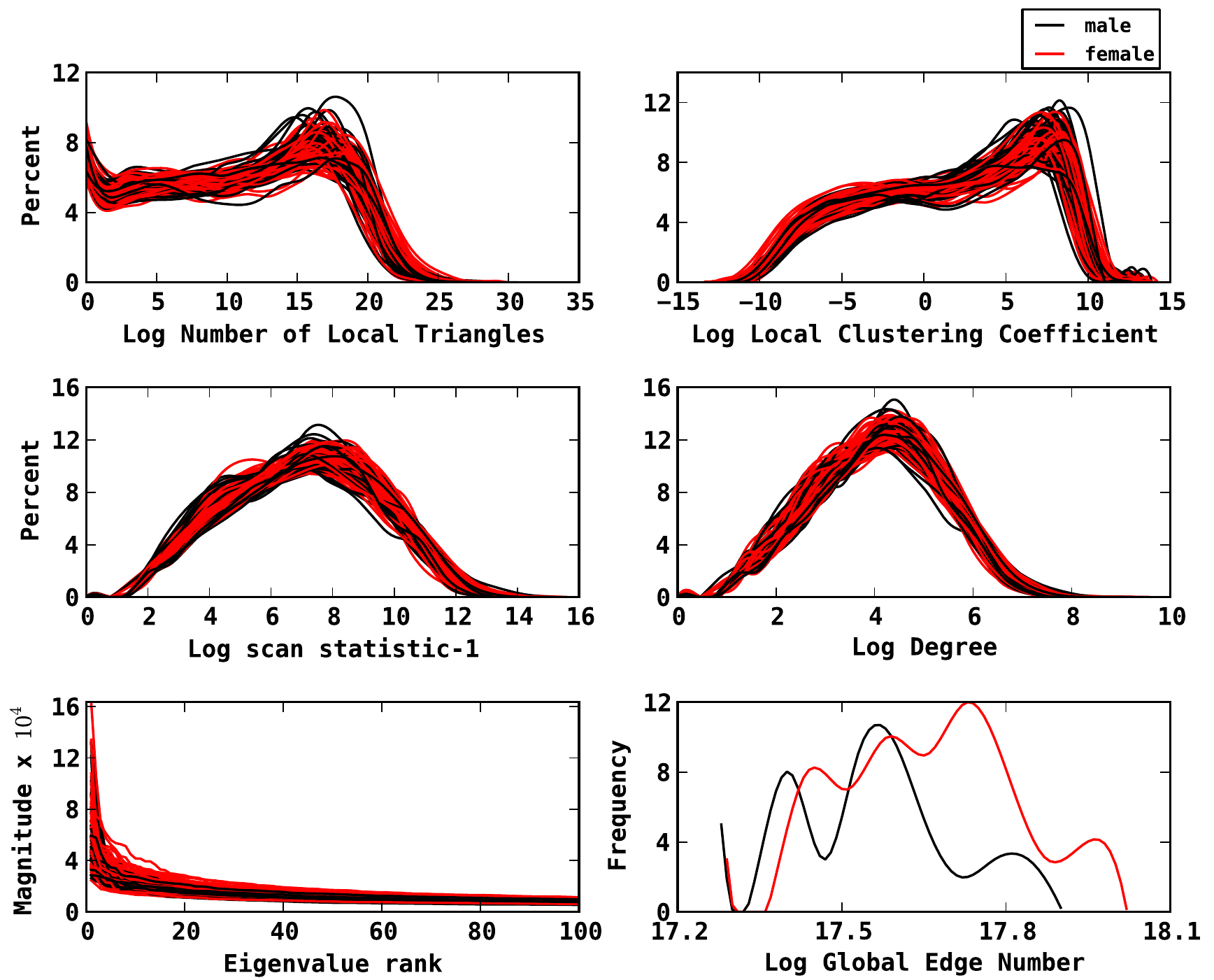}}
\caption{Multivariate invariants of 120 subjects superimposed over one another.} \label{fig:all_inv}

\vspace{-15pt}
\end{figure}

By extrapolating the measured empirical accuracy of smaller graphs, we estimate that given 100 eigen-pairs, the NL-3 algorithm has an accuracy of nearly 94\% on graphs of the LCC containing $O(10^5)$ vertices and $O(10^7)$ edges. Note that Deg and SS-1 are unaffected by the number of eigen-pairs, whereas NL-3 and subsequently CC are affected. Figure \ref{fig:all_inv} depicts all invariants computed on all subjects, with color indicating gender. 
This is an example of how our tools might be used: the invariants may be used as features to discover individual brain differences as a function of phenotype, including psychiatric diagnosis and aptitude.

\section{Computing Performance}
\begin{figure}[!ht]
\vspace{-5pt}
\centering \resizebox{0.5\textwidth}{!}{\includegraphics{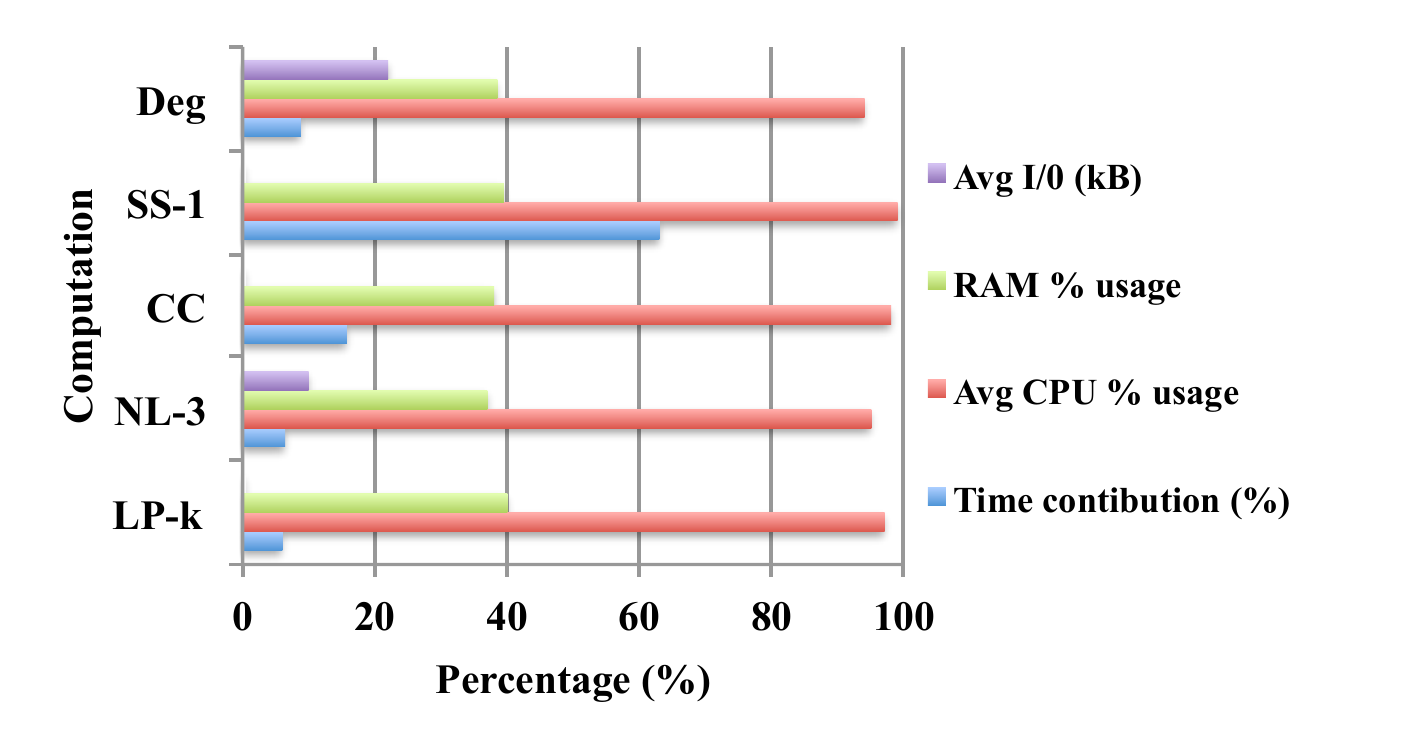}}
\vspace{-17pt}
\caption{ Performance of each invariant, computed serially and independently of any non-dependent invariants on an 4 core, 2.4 GHz processor Linux server with 16 GB of RAM.} \label{fig:time_graph}
\vspace{-13pt}
\end{figure}

To quantify performance we averaged measures such as I/O rates, RAM/CPU usage, for the 120 graphs that we processed (Figure \ref{fig:all_inv}). One of the goals of the computational package is that it be compatible with hardware constraints of modern laptops. Thus, we performed all experiments on a single core, never exceeding 6.5 GB of RAM usage (see Figure \ref{fig:time_graph}). 

For computational efficiency, our implementations are a close-knit merger and coalesced series of invariants. For example, our CC estimate uses our NL-3 estimate which uses our eigendecomposition.  Thus, we compute the collection of multivariate glocal invariants both daisy-chaining appropriate functions and performing each independently.   To facilitate benchmarking against other graph processing packages, Table \ref{tab:time_complex} depicts the computational load for each invariant.  

{Total compute time for all invariants is $\sim$\textbf{2.9} hour per graph, per core, when computed separately. Note that we include the time taken to compute the invariant in addition to any prerequisite computations necessary. Total compute time upon appropriately daisy-chaining is $\sim$\textbf{1.8} h per graph, per core, a speedup of approximately $60\%$.}

\begin{table}
\caption{\small{\textsc{Analysis of algorithms used and experimental time (mean and standard deviation per vertex per core) using a single 4 core, 2.4 GHz processor Linux server with 16 GB of ram. }}}\label{tab:time_complex}
\begin{centering}
\begin{tabular}{ccc}
  Invariant   					& Time Complexity  		& Wall Time  \\
  
  Clustering Coefficient			& $O(n + k)$ 			& $0.59$ $(\pm2.4)$ $\mu$s  \\
 Number of Local 3-Cliques 			& $O(n + k)$ 			& $48.51$ $(\pm0.9)$ $\mu$s \\ 
 Degree 							& $O(n)$ 				& $66.64$ $(\pm1.8)$ $\mu$s \\ 
 Scan Statistic-1 				& $O(nm)$ 				& $0.47$ $(\pm0.2)$ $m$s \\ 
 Latent Position-100								& $O(100(m + n))$			& $45.41$ $(\pm0.1)$ $\mu$s \\

\end{tabular}
\end{centering}
\vspace{-15pt}
\end{table}

\section{Future Work}
We have already incorporated some of this functionality into igraph \cite{Csardi2006} and MIGRAINE \cite{Roncal2013a}, with plans to also incorporate it into neuroimaging specific packages such as  the Configurable Pipeline for the Analysis of Connectomes (CPAC) \cite{CPAC}.

Moreover, we will extend the back-end to take advantage of faster/multithreaded eigensolvers, including GPU implementations.
In order to maximize utility to researchers, we will continue to develop our software suite by computing other network-elucidating invariants %
such as Scan Statistic-2 \cite{Pao2011}, graph diameter, clique number, and more.  Finally, we will utilize these invariants to develop biomarkers for a variety of neuropsychiatric and psychological conditions \cite{APA}.

\section{Acknowledgments}
{\small
This work was supported by the following institutions and grants:
 The National Institutes of Health (NIBIB 1RO1EB016411-01),
 The National Science Foundation (OCI-1244820),
 The Office of Naval Research, and 
 The Johns Hopkins University, Institute for Data Intensive Engineering and Science.
}

\bibliographystyle{nature}
{\small
\bibliography{Invariants,bib2}
}

\end{document}